\documentclass{aa}

\usepackage{graphics}
\usepackage{rotating}
\usepackage{times}

\begin{document}

\thesaurus{05(10.15.2: Praesepe; 08.02.4)}

\title{Investigation of the Praesepe cluster\thanks{based on observations
collected at the Haute-Provence Observatory (France)}}

\subtitle{III. Radial velocity and binarity of the F5-K0 Klein-Wassink stars}

\author{J.-C.~Mermilliod\inst{1} \and M.~Mayor\inst{2}}

\mail{Jean-Claude.Mermilliod@obs.unige.ch}

\institute{Institut d'Astronomie de l'Universit\'e de Lausanne, 
    CH-1290 Chavannes-des-Bois, Switzerland
\and Observatoire de Gen\`{e}ve, CH-1290 Sauverny, Switzerland}

\date{Received date; accepted date}

\maketitle

\begin{abstract}
Coravel observations of 103 F5-K0 stars in the Praesepe cluster yielded
24 spectroscopic binaries (3 are non-members), and 20 orbits were determined, 
with periods from 4 to 7400 days. Based on a complete sample in the colour range 
$0.40 \leq B-V \geq 0.80$ (80 stars, including KW 244 = TX Cnc), the companion 
star fraction CSF = 0.45.
The percentage of spectroscopic binaries with P $<$ 1000$\fd$ is 20\% (16/80).
The combined photometric and spectroscopic analysis showed that 12 among 18 
single-lined spectroscopic binaries are located within the \lq\lq single" star 
sequence in the $(V,B-V)$ diagram and cannot be detected by the photometric 
analysis in the UBV system. In addition, seven photometrically analysed binaries 
were not detected with the radial velocity observations, but are confirmed 
members. The number of single:binary:triple stars is 47:30:3.


\end{abstract}

\section{Introduction}
Observations of star-forming regions (SFR) have demonstrated that stars form mainly 
in clusters and associations, and have a large number of companions. Numerical 
simulations have also shown that the presence of primordial binaries influences 
the dynamical evolution of star clusters. Therefore binarity is a basic information 
to describe properly the results of star-formation processes and the stellar 
population in open clusters. 
Technical developments in the domain of radial-velocity scanner, speckle 
interferometer and adaptive optics imager have permitted to obtain in the 
recent years a wealth of interesting observations which shed new light on the old, 
and so far not solved, question of the similarity, or difference, of the binary 
frequency in star clusters.

Duch\^ene (\cite{gd99}) has reexamined the evidence from SFR and concluded that 
four SFR have a larger rate of binaries than open clusters. However, both kinds 
of stellar systems are not compared exactly in the same period or separation 
intervals. Most recent observations in SFR were performed with speckle interferometry 
or adaptive-optics imaging (see Duch\^ene (\cite{gd99}) for references). 
Among the nearby open clusters, only the Hyades (Mason \cite{mah}; Patience 
et al.~\cite{pgr}) and Pleiades (Bouvier et al.~\cite{brn}) have been observed 
to detect binaries in the range of separation 0$\farcs$05 - 0$\farcs$50. 
The speckle interferometry survey of Mason et al. (\cite{mha}) hardly reached 
the brighter F5 stars and there is little overlap with the sample discussed in 
this paper. Conversely, long term radial-velocity surveys have been undertaken 
(Mermilliod~\cite{jcm97}; Stefanik \& Latham~\cite{sl92}) to monitor main-sequence 
late-type stars in nearby open clusters, while very few data have been published 
for star-forming regions.

This paper is the third one in this series devoted to the investigation of
Praesepe (NGC 2632, M44, $\alpha$ (B1950) = 8$^h$ 37$\fm$2, $\delta$ 
(B1950) = +20$^{\degr}$ 10). In the first paper (Mermilliod et al.~\cite{mwdm}) we
identified 48 new members in the cluster corona, out to 4 degrees and found
10 spectroscopic binaries. Six orbits were determined. In the second paper
(Mermilliod et al.~\cite{mdm}) we presented the orbital elements of three 
spectroscopic binaries in triple systems (\object{KW 365}, \object{KW 367} and 
\object{KW 495}). All the 
material has been used by Raboud \& Mermilliod (\cite{rmb}) to study the radial
structure of Praesepe. It has been shown that the spectroscopic binaries 
are slightly more concentrated toward the cluster centre than single stars. 
Furthermore, they found that the mass function of the primary of spectroscopic 
binaries is different from that for the single stars, confirming a result also 
found in the Pleiades (Raboud \& Mermilliod~\cite{rma}).

Few radial velocities for solar-type dwarfs in Praesepe have been published so 
far. Bolte (\cite{bol}) observed 14 stars to check the binarity status of stars that 
appeared as photometric double stars in the colour-magnitude diagram, having in mind 
the interpretation of the second-sequence stars in globular clusters. Barrado y 
Navascu\'es et al. (\cite{bsr}) have obtained a number of radial velocities (mostly 
one per star) for a sample of G, K and M stars in Praesepe to study the stellar
activity. The agreement of their radial velocities with our measurements is very good.
Very recently, Abt \& Willmarth (\cite{aw99}) published radial-velocity observations of 16
A-type stars on the upper main sequence and determined new orbits for 5 stars.

The sample and the observations are described in sect. 2, the results and orbits 
are presented in sect. 3. Binarity  in Praesepe is discussed in sect. 4.

\section{Observations}
\subsection{The sample}
The initial observing programme included all 88 known F5 - K0 members in the region 
studied by Klein-Wassink (\cite{kw}). The original limiting magnitude was $B$ = 12.5.
In 1977, the membership estimates were mainly based on the proper motions from 
Klein-Wassink (\cite{kw}) and the $UBV$ photometry of Johnson (\cite{jhl}).

Three F5 V stars (\object{KW 295}, \object{KW 478}, \object{KW 555}) did no produce any 
correlation dip. Although their rotational velocities ($V \sin i$) are not known, they 
are probably larger than about 45 km s$^{-1}$, as is frequent for early-F stars. 
In addition, no radial velocities are available for these stars. \object{KW 244} 
(TX Cnc, $V$ = 10.02, $B-V$ = 0.62) is a short period eclipsing binary (Whelan et al. 
\cite{wwm}). It could not be observed with the Coravel because of its rotation, but will 
be included in the binary statistics presented below.

\subsection{Coravel observations}
The observations were obtained with the CORAVEL radial-velocity scanner 
(Baranne et al.~\cite{bmp}) installed on the Swiss 1-m telescope at the 
Haute-Provence Observatory (OHP) for stars later than spectral type F5 and
brighter than $B$ = 12.5. Between November 1978 and December 1996, four to nine 
observations per star were obtained. Binaries were observed more often to derive 
orbital elements. From January 1993 on, 16 fainter stars (B $<$ 13.5) 
were added to the sample and observed, mainly for the purpose of 
determining rotational velocities. One or two observations have been 
obtained. They are listed in Table~\ref{mem} for sake of completeness, because 
few radial velocities have been so far published for these faint members.

The OHP radial velocities were corrected for zero-point differences
to place them in the system defined by Mayor and Maurice (\cite{mama}). The
integration times ranged from about 200 to as much as 1500 seconds,
the average being about 400 seconds. The errors on well-exposed individual 
measurements usually were smaller than 0.5 km s$^{-1}$. However, for a few 
stars with large rotation ($v\sin i \simeq$ 30 km s$^{-1}$) the errors may 
reach 2 km s$^{-1}$, depending on the width of the correlation function. 
The mean value of $\sigma (O-C)$ for the 18 binaries is 0.73 km s$^{-1}$. 

The KW numbers (Klein-Wassink~\cite{kw}), the $V$ magnitudes, the mean radial 
velocities $V_{r}$, the standard errors $\epsilon$ (in km s$^{-1}$), the number 
of measurements $n$, the time intervals $\Delta T$ covered by the observations, 
the probability $P(\chi^2)$ that the scatter is due to chance (Mermilliod \& 
Mayor~\cite{mm89}) are collected in Table~\ref{mem}. The remarks SB1O or SB2O refer to 
single-lined and double-lined binaries for which orbits have been determined, 
PHB to binaries detected photometrically in the colour-magnitude diagram, 
VB to visual binaries. 

\begin{table}[tp]
\caption[]{CORAVEL results for KW stars in Praesepe}
\label{mem}
\begin{flushleft}
\begin{tabular}{rrrrrrrrl}
\hline\noalign{\smallskip}
KW & $m_{V}$ &  \multicolumn{1}{c}{$V_{r}$} & \multicolumn{1}{c}{$\epsilon$} &  \multicolumn{1}{c}{E/I} & $n$ & $\Delta T$ & $P(\chi^2)$ & Rem \\
\hline\noalign{\smallskip}
  9 & 11.39 & 34.85 & 0.35 &  1.94 &  5 & 3986 & 0.005 & PHB \\ 
 16 &  9.18 & 34.33 & 1.71 &  5.56 & 22 & 5858 & 0.000 & SB2 \\ 
 23 & 11.29 & 34.98 & 0.23 &  1.20 &  5 & 5793 & 0.224 &    \\ 
 27 & 11.44 & 34.69 & 0.20 &  1.09 &  5 & 5793 & 0.327 &    \\ 
 30 & 11.40 & 35.45 & 0.23 &  0.63 &  4 & 4458 & 0.756 &    \\ 
 31 &  9.79 & 34.99 & 0.23 &  1.36 &  8 & 5504 & 0.076 & PHB \\ 
 32 & 11.65 & 35.53 & 0.27 &  0.69 &  3 & 2977 & 0.621 &    \\ 
 47 &  9.87 & 34.94 & 0.13 & 21.93 & 34 & 4058 & 0.000 & SB1O \\ 
 49 & 10.66 & 34.71 & 0.28 &  1.29 &  5 & 6594 & 0.156 &    \\ 
 55 & 11.41 & 39.94 & 0.10 & 65.47 & 32 & 3256 & 0.000 & NM \\ 
 58 & 11.26 & 34.35 & 0.22 &  1.12 &  5 & 6175 & 0.291 &    \\ 
 70 & 11.84 & 34.84 & 0.18 &  1.11 &  6 & 6175 & 0.292 &    \\ 
 90 & 10.84 & 35.72 & 0.20 &  1.23 &  7 & 6175 & 0.178 & PHB \\ 
100 & 10.55 & 33.89 & 0.21 &  0.98 &  7 & 7263 & 0.462 &    \\ 
127 & 10.80 & 34.31 & 0.12 & 26.32 & 33 & 5848 & 0.000 & SB1O \\ 
142 &  9.29 & 35.00 & 0.23 & 21.18 & 55 & 5557 & 0.000 & SB2O \\ 
155 &  9.41 & 34.86 & 0.44 &  1.05 & 11 & 4762 & 0.382 &    \\ 
162 & 10.58 & 34.90 & 0.40 &  1.64 &  5 & 6594 & 0.031 &    \\ 
164 & 11.31 & 34.35 & 0.29 &  1.27 &  4 & 4458 & 0.184 &    \\ 
181 & 10.53 & 34.49 & 0.10 & 59.08 & 44 & 5853 & 0.000 & SB1O \\ 
182 & 10.35 & 34.17 & 0.15 &  0.88 &  9 & 7272 & 0.631 & PHB \\ 
184 & 11.56 & 34.95 & 0.11 & 32.13 & 39 & 5049 & 0.000 & SB2O \\ 
196 & 10.74 & 34.99 & 0.23 &  1.01 &  5 & 5173 & 0.398 &    \\ 
208 & 10.66 & 34.12 & 0.29 &  1.29 &  5 & 6594 & 0.162 &    \\ 
213 & 11.81 & 34.34 & 0.21 &  0.97 &  4 & 4460 & 0.418 &    \\ 
217 & 10.23 & 33.64 & 0.33 &  1.63 &  8 & 6976 & 0.008 &    \\ 
218 &  9.34 & 36.74 & 1.22 &  1.14 &  4 & 4024 & 0.276 &    \\ 
222 & 10.11 & 34.29 & 0.24 &  1.19 &  7 & 6965 & 0.209 &    \\ 
227 &  9.49 & 34.77 & 0.41 &  1.64 &  7 & 6532 & 0.013 &    \\ 
238 & 10.29 & 34.28 & 0.36 &  1.13 & 10 & 6965 & 0.074 &    \\ 
239 &  9.66 & 36.66 & 0.54 &  1.12 &  8 & 4760 & 0.292 &    \\ 
250 &  9.79 & 36.45 & 0.74 &  1.00 &  6 & 2509 & 0.425 &    \\ 
257 & 11.01 & 35.13 & 0.44 &  2.12 &  6 & 7230 & 0.000 & PHB \\ 
258 & 10.24 & 54.69 & 0.11 & 14.62 & 31 & 3724 & 0.000 & NM \\ 
268 &  9.89 & 35.75 & 0.13 & 13.34 & 44 & 4459 & 0.000 & SB1O \\ 
275 &  9.96 & 34.45 & 0.17 &  1.35 & 10 & 7272 & 0.060 & PHB \\ 
287 & 10.37 & 34.09 & 0.10 &  5.31 & 28 & 7207 & 0.000 & SB1O \\ 
288 & 10.70 & 35.32 & 0.25 &  1.32 &  6 & 6976 & 0.132 &    \\ 
293 &  9.85 & 36.61 & 0.56 &  0.89 &  7 & 3670 & 0.574 &    \\ 
297 & 11.64 & 37.14 & 0.27 &  2.43 & 16 & 6118 & 0.000 & SB1 \\ 
301 & 11.17 & 34.11 & 0.35 &  1.72 &  5 & 6175 & 0.020 &    \\ 
304 & 11.52 & 33.53 & 0.26 &  0.44 &  4 & 4457 & 0.903 &    \\ 
309 & 11.63 & 34.42 & 0.21 &  0.66 &  4 & 3668 & 0.729 &    \\ 
322 & 10.87 & 35.08 & 0.45 &  3.91 & 17 & 7228 & 0.000 & SB1 \\ 
325 & 10.60 & 34.63 & 0.11 & 12.40 & 30 & 6139 & 0.000 & SB1O \\ 
326 & 11.35 & 34.28 & 0.19 &  0.89 &  5 & 3665 & 0.535 &    \\ 
332 &  9.56 & 33.44 & 0.44 &  0.89 &  8 & 6259 & 0.597 &    \\ 
334 & 11.00 & 36.35 & 0.32 &  1.78 &  7 & 7230 & 0.005 & PHB \\ 
335 & 11.02 & 34.62 & 0.23 &  1.15 &  5 & 5793 & 0.264 &    \\ 
336 & 11.46 & 35.21 & 0.25 &  1.11 &  4 & 3665 & 0.294 &    \\ 
341 & 10.30 & 36.13 & 0.38 &  2.00 &  8 & 6177 & 0.000 & SB1 \\ 
365 & 10.18 & 27.21 & 2.31 & 26.64 & 51 & 4728 & 0.000 & Triple \\ 
367 & 10.69 &   var & 8.39 & 42.35 & 82 & 6591 & 0.000 & Triple \\ 
368 & 11.49 & 34.87 & 0.09 & 25.25 & 28 & 4756 & 0.000 & SB1O \\ 
371 & 10.10 & 34.83 & 0.40 &  1.07 &  6 & 3605 & 0.339 & PHB \\ 
392 & 10.76 & 34.57 & 0.25 &  1.43 &  6 & 6176 & 0.072 &    \\ 
\noalign{\smallskip}
\hline
\end{tabular}
\end{flushleft}
\end{table} 

\setcounter{table}{0}
\begin{table}[th!]
\caption[]{(continued)}
\begin{flushleft}
\begin{tabular}{rrrrrrrrl}
\hline\noalign{\smallskip}
KW & $m_{V}$ &  \multicolumn{1}{c}{$V_{r}$} & \multicolumn{1}{c}{$\epsilon$} &  \multicolumn{1}{c}{E/I} & $n$ & $\Delta T$ & $P(\chi^2)$ & Rem \\
\hline\noalign{\smallskip}
396 &  9.85 & 34.15 & 0.34 &  0.83 &  5 & 3607 & 0.604 &    \\ 
399 & 10.97 & 33.32 & 0.37 &  2.30 &  8 & 6175 & 0.000 & SB1 \\ 
403 & 11.71 & 34.39 & 0.37 &  1.85 &  4 & 3667 & 0.016 &    \\ 
411 &  9.32 & 35.63 & 1.42 &  1.66 &  3 & 1037 & 0.063 &    \\ 
416 &  9.59 & 34.49 & 0.11 & 37.84 & 40 & 5115 & 0.000 & SB1O \\ 
418 & 10.47 & 33.87 & 0.27 &  1.45 &  6 & 6176 & 0.066 &    \\ 
421 & 10.15 & 34.87 & 0.22 &  1.00 &  6 & 6177 & 0.418 &    \\ 
425 & 11.42 & 40.48 & 0.32 &  1.52 &  7 & 6174 & 0.035 & NM \\ 
432 & 11.05 & 32.99 & 0.21 &  0.96 &  4 & 5792 & 0.452 &    \\ 
434 & 11.41 & 34.34 & 0.13 & 45.73 & 21 & 3256 & 0.000 & SB1O \\ 
439 &  9.43 & 34.92 & 0.12 &  3.76 & 34 & 6231 & 0.000 & SB1O \\ 
454 &  9.88 & 34.89 & 0.32 &  0.61 &  6 & 6230 & 0.867 &    \\ 
458 &  9.71 & 33.64 & 0.33 &  1.51 &  7 & 7272 & 0.034 & VB, 0$\farcs$2 \\ 
466 & 10.99 & 33.45 & 0.26 &  1.26 &  5 & 6174 & 0.195 &    \\ 
476 & 11.62 & 34.70 & 0.27 &  1.26 &  4 & 3667 & 0.199 &    \\ 
488 & 11.44 & 34.69 & 0.24 &  1.13 &  4 & 3665 & 0.287 &    \\ 
495 &  9.97 & 32.73 & 0.15 &  0.81 & 31 & 5114 & 0.138 & SB3O \\ 
496 &  9.57 & 33.68 & 0.22 &  0.78 & 11 & 6212 & 0.826 & SB2 \\ 
498 & 11.78 & 33.83 & 0.23 &  0.58 &  4 & 3667 & 0.797 &    \\ 
508 & 10.77 & 34.23 & 0.64 &  6.32 & 26 & 6174 & 0.000 & SB1O \\ 
533 & 11.58 & 35.00 & 0.18 &  1.62 & 14 & 6171 & 0.001 & SB2 \\ 
537 & 11.65 & 33.95 & 0.21 &  1.04 &  4 & 3666 & 0.363 &    \\ 
539 & 11.18 & 34.25 & 0.52 &  4.88 & 20 & 6171 & 0.000 & SB1O \\ 
540 & 11.03 & 36.80 & 0.14 &  4.80 & 24 & 6171 & 0.000 & SB1O \\ 
542 & 11.72 & 33.60 & 0.22 &  1.03 &  4 & 3666 & 0.363 &    \\ 
546 & 11.62 & 37.78 & 0.27 &  0.74 &  2 &  684 & 0.457 &    \\ 
548 & 10.01 & -9.14 & 0.15 &  0.95 &  6 & 5081 & 0.483 & NM \\ 
553 & 10.15 & 27.16 & 0.25 & 32.82 & 48 & 5166 & 0.000 & NM \\ 
557 & 10.38 &  4.94 & 0.36 &  1.58 &  4 & 3991 & 0.061 & NM \\ 
\noalign{\medskip}
\noalign{Fainter stars}
\noalign{\medskip}
 48 & 12.32 & 34.69 & 0.37 &       &  1 &      &       &    \\ 
 52 & 12.28 & 35.05 & 0.39 &       &  1 &      &       &    \\ 
 79 & 12.08 & 36.03 & 0.40 &       &  1 &      &       & PHB \\ 
236 & 12.94 & 36.08 & 0.52 &  0.15 &  2 &    2 & 0.878 & Triple? \\ 
246 & 12.01 & 26.83 & 1.18 &  4.27 &  2 &  684 & 0.000 & SB1 \\ 
263 & 12.01 & 34.00 & 0.40 &       &  1 &      &       &    \\ 
313 & 12.20 & 34.80 & 0.37 &       &  1 &      &       &    \\ 
344 & 12.10 & 34.39 & 0.37 &       &  1 &      &       &    \\ 
349 & 12.23 & 34.26 & 0.36 &       &  1 &      &       &    \\ 
353 & 12.35 & 33.85 & 0.39 &       &  1 &      &       &    \\ 
417 & 12.35 & 36.72 & 0.39 &       &  1 &      &       &    \\ 
430 & 12.06 & 35.20 & 0.27 &  0.18 &  2 & 1035 & 0.858 &    \\ 
448 & 12.15 & 34.98 & 0.38 &       &  1 &      &       &    \\ 
471 & 12.15 & 35.07 & 0.35 &       &  1 &      &       &    \\ 
474 & 12.12 & 32.36 & 0.36 &       &  1 &      &       & PHB \\ 
492 & 12.13 & 33.93 & 0.39 &       &  1 &      &       &    \\ 
530 & 12.26 & 77.18 & 0.39 &  1.71 &  4 & 3357 & 0.035 & NM \\ 
547 & 11.99 & 39.25 & 1.25 &  5.99 &  3 & 1035 & 0.000 & SB1 \\ 
\noalign{\smallskip}
\hline
\end{tabular}
\end{flushleft}
\end{table} 

Individual observations can be requested from the first author.
The whole CORAVEL dataset is currently being recalibrated for possible zero-point 
error and color effect. All the individual data will then be published in a 
comprehensive catalogue of observations in Praesepe.

\section{Results}
\subsection{Membership}
The radial-velocity results summarized in Table~\ref{mem} confirm the membership of 
most stars, with seven exceptions. KW 55, 258, 425, 530, 548, 553 and 557 are clearly 
(more than 7 $\sigma$) non-members, although the membership probabilities from proper 
motions (Jones \& Cudworth~\cite{jc}) for \object{KW 55} (P = 0.98), \object{KW 258} 
(P = 0.94) and \object{KW 425} (P = 0.86) are quite high. 
KW 55, KW 258 and \object{KW 553} are spectroscopic binaries, and their systemic velocities 
(+39.94, +54.7 and +27.2 km s$^{-1}$, respectively) differ from the mean cluster velocity 
by at least 7 km$^{-1}$. The scatter of the ($O-C$) residuals doest not support any 
long term variability which would indicate that these stars are in triple systems. 
\object{KW 425}, \object{KW 530} and KW 553 are well below the ZAMS in the colour-magnitude 
diagram, while \object{KW 548} and \object{KW 557} are above the main sequence band.

\subsection{Spectroscopic binaries}
Thirty spectroscopic binaries and triple systems (27 members and 3 non-members) have
been discovered among Praesepe F5 - K0 stars. Twenty orbits have been determined
(17 members, 3 non-members). The shortest period is 3$\fd$93 and the longest 7365$\fd$.
The orbital elements of a few binaries could not be determined either because the 
expected period is longer than twenty years, or because the frequency of observations 
did not permit to reobserve several times the double-lined patterns, the radial 
velocity being otherwise constant at the cluster velocity (KW 16, 496, 533).
None of the stars considered as non variable show any trend with time, only star 
KW 162 shows a slight change from +36.3 km s$^{-1}$ in 1978 to +34.5 km s$^{-1}$ in 
1990, with values of +35.0 and +35.5 in 1982 and 1986. The evidence remains marginal. 
We can conclude that the undiscovered binaries will have radial-velocity amplitude 
smaller than about 2 km s$^{-1}$ and periods longer than, at least, 15 years.

The orbital elements for 20 new orbits are presented in Table~\ref{orbs},
and the in-phase radial-velocity diagrams are presented for each binary. To the
19 KW stars, the elements for \object{VL 1025} have been added. This star is slightly 
outside the Klein-Wassink area and its binarity has been detected in Paper I 
(Mermilliod et al.~\cite{mwdm}). An orbit has now been determined and the 
elements are included at the end of Table~\ref{orbs}.

Orbital elements for the spectroscopic binaries \object{KW 365}, \object{KW 367} and 
\object{KW 495} belonging to triple systems have been published by Mermilliod et al. 
(\cite{mdm}). The periods given in Table~\ref{orbs} for KW 367 correspond to the short 
(Aab) and long period (Aab x B), respectively.

\begin{table*}
\caption{Orbital parameters}
\label{orbs}
\begin{flushleft}
\begin{tabular}{rr@{.}lr@{.}lr@{.}lr@{.}lr@{.}lr@{.}lr@{.}lr@{.}lr@{.}lr@{.}lr}
\noalign{\smallskip}
\hline
\noalign{\smallskip}
 KW  & \multicolumn{2}{c}{Period} & \multicolumn{2}{c}{T} & \multicolumn{2}{c}{e} & \multicolumn{2}{c}{$\gamma$} & 
\multicolumn{2}{c}{$\omega$} & \multicolumn{2}{c}{K$_1$} & \multicolumn{2}{c}{K$_2$} & \multicolumn{2}{c}{f(m)} & 
\multicolumn{2}{c}{a $\sin$ i} & \multicolumn{2}{c}{$\sigma (O-C)$} & n$_{obs}$ \\
\noalign{\smallskip}
\hline
\noalign{\smallskip}
  47 &  34&619    & 8989&69  & 0&406 &   34&94  &   289&8  & 19&01 & \multicolumn{2}{c}{} & 0&0189 &   8&27     &   0&72         &   34  \\
     &    &002    &     &25  &  &009 &     &13  &     1&6  &   &18 & \multicolumn{2}{c}{} &  &0008 &    &11     &  \multicolumn{2}{c}{} &       \\
  55 &   5&98533  & 9002&06  & 0&00  &   39&94  & \multicolumn{2}{c}{} & 34&73 & \multicolumn{2}{c}{} & 0&0260 &   2&858    &   0&57         &   32  \\
     &    &00003  &     &01  &  &004 &     &10  & \multicolumn{2}{c}{} &   &14 & \multicolumn{2}{c}{} &  &0003 &    &011    & \multicolumn{2}{c}{} &       \\
 127 &  13&2803   & 9002&61  & 0&226 &   34&31  &   130&7  & 17&82 & \multicolumn{2}{c}{} & 0&0072 &   3&170    &   0&61         &   33  \\
     &    &0001   &     &08  &  &009 &     &12  &     2&5  &   &19 & \multicolumn{2}{c}{} &  &0003 &    &040    & \multicolumn{2}{c}{} &       \\
 142 &  45&9746   & 9972&36  & 0&223 &   34&90  &    56&7  & 32&50 & 33&92 & 0&1725  &   20&90  &  2&67         &   55  \\
     &    &0019   &     &41  &  &012 &     &28  &     3&3  &   &52 &   &48 &  &0089  &     &35  & \multicolumn{2}{c}{} &       \\
 181 &   5&866276 & 8998&266 & 0&357 &   34&49  &   228&9  & 47&13 & \multicolumn{2}{c}{} & 0&0520 &   3&552    &   0&65         &   44  \\
     &    &000006 &     &006 &  &003 &     &10  &      &5  &   &14 & \multicolumn{2}{c}{} &  &0006 &    &015    & \multicolumn{2}{c}{} &       \\
 184 &  47&438    & 8969&58  & 0&341 &   34&95  &    68&2  & 23&46 & 24&17 & 0&0529 &  14&39     &   0&94         &   39  \\
     &    &002    &     &15  &  &007 &     &11  &     1&3  &   &23 &   &24 &  &0029 &    &18     & \multicolumn{2}{c}{} &       \\
 258 & 260&41     & 4802&6   & 0&394 &   54&69  &   194&9  & 11&95 & \multicolumn{2}{c}{} & 0&0358 &  39&3      &   0&55         &   31  \\
     &    &13     &    1&2   &  &015 &     &11  &     2&2  &   &29 & \multicolumn{2}{c}{} &  &0034 &   1&3      & \multicolumn{2}{c}{} &       \\
 268 & 144&340    & 4838&50  & 0&649 &   35&75  &   285&3  & 13&25 & \multicolumn{2}{c}{} & 0&0153 &  20&00     &   0&67         &   44  \\
     &    &013    &     &33  &  &008 &     &13  &     1&5  &   &20 & \multicolumn{2}{c}{} &  &0011 &    &49     & \multicolumn{2}{c}{} &       \\
 287 &7365&       & 6249&    & 0&62  &   34&09  &    72&9  &  4&86 & \multicolumn{2}{c}{} & 0&0431 & 388&       &   0&46         &   28  \\
     & 325&       &   38&    &  &03  &     &09  &     4&0  &   &17 & \multicolumn{2}{c}{} &  &0099 &  41&       & \multicolumn{2}{c}{} &       \\
 325 & 896&9      & 9164&    & 0&117 &   34&63  &   146&5  &  8&79 & \multicolumn{2}{c}{} & 0&0619 & 107&6      &   0&52         &   30  \\
     &   1&4      &   18&    &  &020 &     &11  &     8&0  &   &15 & \multicolumn{2}{c}{} &  &0038 &   2&3      & \multicolumn{2}{c}{} &       \\
 367 &   3&057311 & 9998&9   & 0&000 &   33&41  & \multicolumn{2}{c}{} & 52&88 & 66&16 & \multicolumn{2}{c}{} & \multicolumn{2}{c}{} & \multicolumn{2}{c}{} & 88  \\
     &    &000002 &    1&9   &  &005 &     &20  & \multicolumn{2}{c}{} &   &14 &   &37 & \multicolumn{2}{c}{} & \multicolumn{2}{c}{} & \multicolumn{2}{c}{}  &     \\
 367 &1659&       & 8025&    & 0&762 &   33&41  &  230&5   &  8&32 & \multicolumn{2}{c}{} & 0&0318 & 136&       & \multicolumn{2}{c}{} &   88  \\
     &  14&       &   10&    &  &024 &     &20  &    4&6   &   &90 & \multicolumn{2}{c}{} & \multicolumn{2}{c}{} & \multicolumn{2}{c}{}  & \multicolumn{2}{c}{} &       \\
 368 &  76&5643   & 9478&18  & 0&212 &   34&87  &    12&8  & 14&53 & \multicolumn{2}{c}{} & 0&02275& 149&47     &   0&47         &   28  \\
     &    &0045   &     &48  &  &008 &     &09  &     2&5  &   &13 & \multicolumn{2}{c}{} &  &00076&    &17     & \multicolumn{2}{c}{} &       \\
 416 &  25&84234  & 9978&406 & 0&432 &   34&49  &   318&88 & 26&94 & \multicolumn{2}{c}{} & 0&03850&   8&635    &   0&60         &   40  \\
     &    &00016  &     &038 &  &005 &     &11  &      &84 &   &16 & \multicolumn{2}{c}{} &  &00098&    &073    & \multicolumn{2}{c}{} &       \\
 434 &   3&932626 & 9998&105 &(0&000)&   34&34  & \multicolumn{2}{c}{} &  33&27 & \multicolumn{2}{c}{} & 0&01504 & 1&7991 &  0&50    &   20  \\
     &    &000012 &     &015 &  &005 &     &13  & \multicolumn{2}{c}{} &    &18 & \multicolumn{2}{c}{} &  &00025 &  &0098 & \multicolumn{2}{c}{} &       \\
 439 & 457&8      & 9332&    & 0&17  &   34&92  &    34&   &  2&62 & \multicolumn{2}{c}{}  & 0&00082&  16&2      &   0&60         &   34  \\
     &   1&4      &   27&    &  &06  &     &12  &    23&   &   &19 & \multicolumn{2}{c}{} &  &00021&   1&4      & \multicolumn{2}{c}{} &       \\
 508 & 647&64     & 9477&1   & 0&496 &   34&19  &   171&1  &  4&72 & \multicolumn{2}{c}{} & 0&0046 &  36&5      &   0&55         &   26  \\
     &    &77     &    5&7   &  &039 &     &13  &     5&0  &   &28 & \multicolumn{2}{c}{} &  &0012 &   3&2      & \multicolumn{2}{c}{} &       \\
 539 &5551&       & 6966&    & 0&533 &   34&03  &    49&1  &  4&04 & \multicolumn{2}{c}{} & 0&0236 & 263&       &   0&42         &   20  \\
     & 221&       &   94&    &  &049 &     &18  &     6&8  &   &25 & \multicolumn{2}{c}{} &  &0079 &  36&       & \multicolumn{2}{c}{} &       \\
 540 &1149&5      & 9082&    & 0&669 &   36&75  &   356&0  &  4&05 & \multicolumn{2}{c}{} & 0&0033 &  48&       &   0&53         &   24  \\
     &   6&5      &   12&    &  &080 &     &13  &     5&5  &   &91 & \multicolumn{2}{c}{} &  &0032 &  16&       & \multicolumn{2}{c}{} &       \\
 553 &  40&6649   & 9969&23  & 0&342 &   27&16  &   266&2  & 47&47 & \multicolumn{2}{c}{} & 0&375  &  24&95     &   1&57         &   47  \\
     &    &0014   &     &11  &  &008 &     &25  &     1&2  &   &33 & \multicolumn{2}{c}{} &  &011  &    &25     & \multicolumn{2}{c}{} &       \\
VL 1025 &2657&       & 7203&    & 0&481 &   33&95  &   275&2  &  6&22 & \multicolumn{2}{c}{} & 0&0448 & 199&       &   0&52         &   22  \\
     &  26&       &   18&    &  &029 &     &11  &     4&3  &   &21 & \multicolumn{2}{c}{} &  &0075 &  12&       & \multicolumn{2}{c}{} &       \\
\noalign{\smallskip}
\hline
\end{tabular}
\end{flushleft}
\end{table*}

\begin{figure}
\resizebox{7.0cm}{!}{\includegraphics{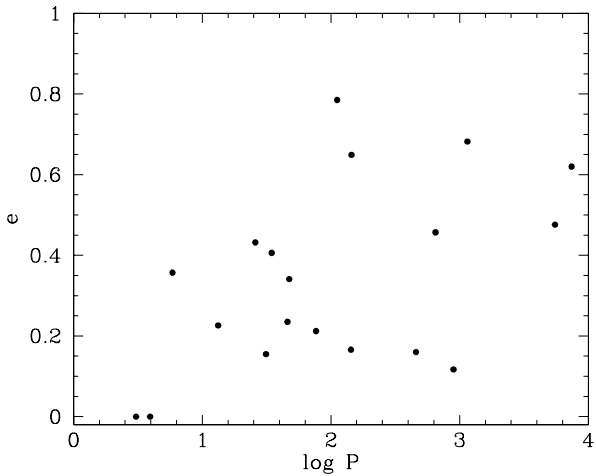}}
\caption{Eccentricity - period diagram.}
\label{elogp}
\end{figure}

The diagram of the eccentricity vs logarithm of the periods (Fig.~\ref{elogp}) shows 
the clear transition from circular to elliptical orbits at about 8 days. \object{KW 181} 
(P = 5$\fd$86, e = 0.36) is an exception. There is so far no indication that this star 
is a triple system. Its original eccentricity must have been quite large, and it had not 
enough time to reduce it to 0 (Duquennoy et al.~\cite{dmm}).

\subsection{Individual stars}

\noindent {\it \object{KW 16}}:
Double lines have been observed on December 14, 1993, with a velocity difference 
of 43 km s$^{-1}$. The 21 velocities of the blend vary between +28 and +43 km s$^{-1}$,
but no period could be determined. The width of the correlation function varies also. 
The double-lined status is in agreement with the position of the star in the CMD.

\noindent {\it \object{KW 236}}:
If it is a member, this star is probably a triple system: it shows a dip at the cluster
velocity and a well marked secondary dip.

\noindent {\it \object{KW 287}}:
This binary has quite a long period, and the first cycle has not yet been 
completely covered. However, a reasonably well determined orbit has been 
computed, because both extrema have been well observed.

\noindent {\it \object{KW 297}}:
The null probability P($\chi^2$) of this star is mostly due to the first
velocity in January 1981 (34.4 km s$^{-1}$). But the present mean velocity
is around 37.5 km s$^{-1}$, which deviates from the cluster mean. It is
probably a very long period binary.

\noindent {\it \object{KW 322}}:
A very preliminary orbit has been obtained, with a period of about 10000 days. 
One complete cycle has not been yet covered and the uncertainties on the elements 
are therefore rather large. The elements are not included in Table~\ref{orbs}

\begin{figure}
\begin{turn}{-90}
\resizebox{6.0cm}{!}{\includegraphics{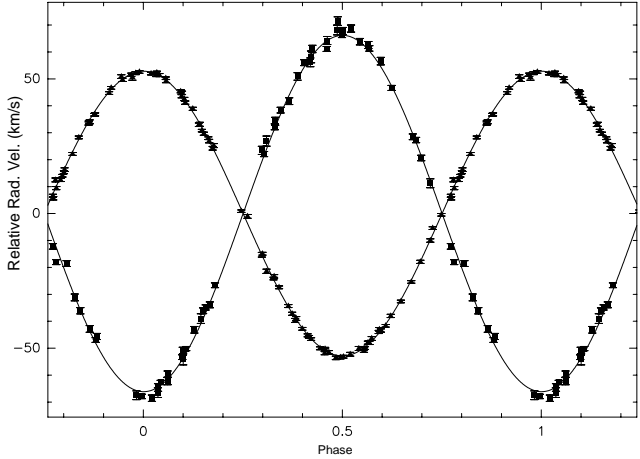}}
\end{turn}
\caption{Radial-velocity curves for the SB2 system in KW 367.}
\label{kws}
\end{figure}

\begin{figure}
\begin{turn}{-90}
\resizebox{6.0cm}{!}{\includegraphics{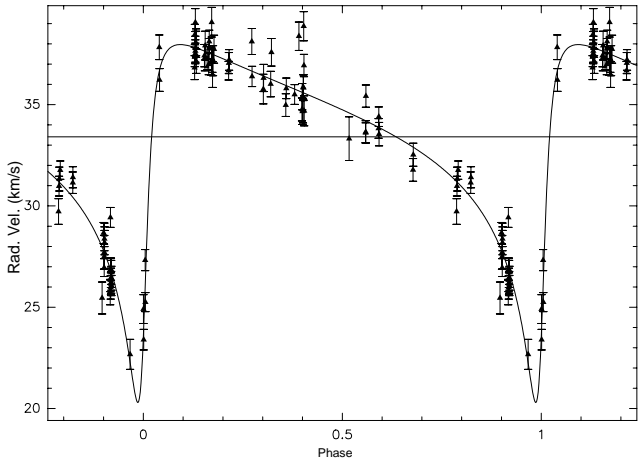}}
\end{turn}
\caption{In phase systemic-velocity variation for the SB2 in KW 367.} 
\label{kwl}
\end{figure}

\noindent {\it \object{KW 367}}:
This is a double-lined binary for which an orbit has already been determined 
(Mermilliod et al. 1994) and the systemic velocity was found to be variable. 
This star has been frequently observed and an orbit for the long period system 
has been determined. The elements are included in Table~2.
The two periods are thus 3$\fd$057 for the close system and 1659$\fd$ for
the longer one. The elements for the short periods are very similar to the first
published ones. The present values has been obtained with additional observations and 
a simultaneous solution for the two orbits. The radial-velocity curves are displayed 
in Figs.~\ref{kws} and~\ref{kwl}. The third component has not been convincingly  
seen in the correlation functions, although a few dips, similar to those of the 
secondary of the SB2, but with completely different velocities could be tentatively 
attributed to the secondary of the long system. If this fact is confirmed, the secondary
could also be a spectroscopic binary itself. However, as deduced from the CM diagram, 
most of the light comes from the SB2 and the other system does not seem to contribute 
much light.

\noindent {\it \object{KW 399}}:
The peak to peak amplitude is 2.68 km s$^{-1}$ and is due to two low velocities
obtained in February 1982 and January 1983. The last observation was made in 1991 
and more recent data are necessary to judge the binary status of this star.

\noindent {\it \object{KW 496}}:
Line doubling has been observed on three nights in 1981, 1983 and 1990, but
the velocity is stable for the 30 other observations. The velocity separation is
close to 20 km s$^{-1}$.

\noindent {\it \object{KW 533}}:
Line doubling has been observed on January and February 1987 and on February 1988, 
with a velocity separation of the order of 25 km s$^{-1}$. Since that time, the 
velocity is constant at the cluster mean velocity. This is probably another 
long period binary.

\subsection{Binary frequency}
We limit our analysis to the stars listed in the first part of Table~\ref{mem} which
corresponds to the colour interval 0.40 $< \,B-V\,<$ 0.80. All stars have been observed
at least 4 times, and the star census is supposed to be complete. We count 80 stars,
including KW 244 (TX Cnc). We find 16 single-lined spectroscopic binaries (SB1), 5 double-lined
binaries (SB2), including \object{KW 244}, 8 photometrically analysed binaries (PHB: 
star well above the ZAMS, $\delta V\,>$ 0.45 mag, but not detected as spectroscopic binaries), 
3 triple systems, one visual binary.

The overall rate in this colour interval is 30\% (24/80) and that for the Pleiades was 19\%
(17/88). Due to the size of the samples, the statistical {\it a posteriori} significance of 
this difference is at the 10\% level. The rate of binaries with 
periods shorter than 1000$^d$ is 20\% (16/80), by counting 
the short-period spectroscopic binaries in the 3 triple systems. This value is slightly 
larger than in the Pleiades (13\%: 11/88) in a similar colour interval, and is significant 
at the 18\% level, and than in the nearby G dwarf sample of Duquennoy \& Mayor 
(\cite{dm91}): 13\% (21/164), with a significance at the level of 14\%. 
In Praesepe, 67\% (16/24) of the binaries have periods shorter than 1000$^d$. The 
longest period determined from our material is 7365 days. This means that the detection 
of binaries with periods shorter should be quite complete.
The proportion of single:binary:triple systems is 47:30:3 for Praesepe. That obtained
for the Pleiades in the range 0.40 $< \,B-V\,<$ 0.90, i.e. slightly larger, was: 56:30:2. 

The values of the multiple-system fraction is $MSF = (b + t) / (s + b + t)$ = 0.41 (33/80) 
and the companion star fraction, $CSF = (b + 2t) / (s + b + t)$ = 0.45 (36/80), 
where $s$ means single, $b$, binary and $t$, triple. 
For a sample of 162 stars in the Hyades, Patience et al. (\cite{pgr}) obtained 
$MSF$ = 0.41 $\pm$ 0.05 and $CSF$ = 0.46 $\pm$ 0.05.
Taking into account that a more extensive binary search has been performed
in the Hyades, the agreement is pretty good, and shows that the multiplicity rates
are not very different between the Hyades and Praesepe. At least, it cannot explain
the striking difference in the X-ray source detection between the two clusters
(Randich \& Schmitt~\cite{rs}, Barrado y Navascu\'es et al.~\cite{bsr}).

\subsection{Mass ratios}
Mass ratios have been computed for the spectroscopic and photometric binaries in
various ways. For photometric binaries, a photometric deconvolution has been 
attempted. The resulting values are given in Table~\ref{ratio}: $V_a$, $C_a$, $M_a$
are $V$ mag., $B-V$ colour and mass of the primary respectively, 
and $V_b$, $C_b$, $M_a$ are the $V$ mag., $B-V$ colour and mass of the secondary.
When the stars were at the upper binary limit or slightly above, 
$V_a$ = $V_{ab}$ - 0.75 mass ratios have been set at 0.99. For spectroscopic 
binaries located well within the single star sequence, the minimum mass (Min) 
has been computed from the spectroscopic orbit and the maximum mass (Max), under 
the assumption that the secondary is 5 mag. fainter than the primary. This limit
corresponds to an effect of 0.01 mag on the V magnitude. A difference of 4 mag. 
may appear more appropriate. In this case, the maximum mass would be larger
and the mass interval for the secondary would be wider. The two 
mass values give a reasonnable estimate of the secondary masses. Values of 
the mass ratios $q$ tabulated in Table~\ref{ratio} show that photometric
detection produces mass ratios larger than 0.60, while the lower limit of the 
mass ratios for the single-lined spectroscopic binaries is around 0.20.

The distribution of $q$, by bins of 0.2 M$_{\odot}$ is 0.8 - 1: 12, 0.6 - 0.8: 9,
0.4 - 0.6: 7, 0.2 - 0.4: 7. There is a slight increase toward values of $q$ larger
than 0.7, but the distribution for $q$ $<$ 0.5 may be incomplete, because the
orbits with very long periods have not been determined yet.
 
\begin{table}
\caption{Deduced mass ratios}
\label{ratio}
\begin{flushleft}
\begin{tabular}{rrrrrrrrrr}
\noalign{\smallskip}
\hline
\noalign{\smallskip}
 KW & \multicolumn{1}{c}{V$_a$} &\multicolumn{1}{c}{V$_b$} & \multicolumn{1}{c}{C$_a$} & 
\multicolumn{1}{c}{C$_b$} & M$_a$ & M$_b$ & \multicolumn{1}{c}{q} & Max & Min \\
\noalign{\smallskip}
\hline
\noalign{\smallskip}
  9 & 11.60 & 13.27 & 0.75 & 1.09 & 0.95 & 0.73 & 0.77 &      &      \\
 16 &  9.72 & 10.29 & 0.43 & 0.51 & 1.27 & 1.16 & 0.92 &      &      \\
 31 & 10.54 &       & 0.56 &      & 1.11 &      & 0.99 &      &      \\
 47 &  9.87 &       & 0.48 &      & 1.22 &      & $>$0.29 & 0.58 & 0.36 \\
 90 & 11.10 & 12.77 & 0.65 & 0.98 & 1.03 & 0.78 & 0.76 &      &      \\
127 & 10.81 &       & 0.60 &      & 1.07 &      & $>$0.23 & 0.40 & 0.23 \\
142 & 10.06 &       & 0.49 &      & 1.19 &      & 0.99 &      &      \\
181 & 10.57 & 13.61 & 0.55 & 1.18 & 1.12 & 0.68 & 0.61 &      & 0.51 \\
182 & 10.97 & 11.22 & 0.63 & 0.68 & 1.04 & 1.00 & 0.96 &      &      \\
184 & 12.38 &       & 0.90 &      & 0.83 &      & 0.99 &      &      \\
268 &  9.89 &       & 0.47 &      & 1.23 &      & $>$0.27 & 0.58 & 0.33 \\
275 & 10.66 & 10.78 & 0.57 & 0.59 & 1.10 & 1.08 & 0.98 &      &      \\
287 & 10.54 & 12.49 & 0.54 & 0.92 & 1.12 & 0.82 & 0.73 &      & 0.48 \\
297 & 11.88 & 13.38 & 0.80 & 1.12 & 0.90 & 0.71 & 0.79 &      &      \\
322 & 11.03 & 13.05 & 0.63 & 1.05 & 1.04 & 0 75 & 0.72 &      &      \\
325 & 10.69 & 13.33 & 0.58 & 1.10 & 1.09 & 0.72 & 0.66 &      & 0.55 \\
334 & 12.20 & 12.98 & 0.67 & 1.04 & 1.01 & 0.75 & 0.75 &      &      \\
367 & 11.02 & 12.17 & 0.63 & 0.85 & 1.04 & 0.87 & 0.83 &      &      \\
368 & 11.51 &       & 0.73 &      & 0.95 &      & 0.35 & 0.33 & 0.33 \\
399 & 10.95 &       & 0.63 &      & 1.05 &      &      & 0.40 &      \\
416 &  9.59 &       & 0.41 &      & 1.31 &      & $>$0.38 & 0.59 & 0.50 \\
434 & 11.39 &       & 0.71 &      & 0.97 &      & $>$0.29 & 0.35 & 0.28 \\
439 &  9.45 &       & 0.40 &      & 1.34 &      & $>$0.09 & 0.60 & 0.12 \\
458 & 10.47 &       & 0.55 &      & 1.12 &      & 0.99 &      &      \\
496 & 10.09 & 10.71 & 0.48 & 0.59 & 1.20 & 1.08 & 0.90 &      &      \\
508 & 10.78 &       & 0.59 &      & 1.08 &      & $>$0.18 & 0.25 & 0.19 \\
533 & 12.34 &       & 0.90 &      & 0.83 &      & 0.99 &      &      \\
540 & 11.13 & 13.71 & 0.65 & 1.21 & 1.02 & 0.67 & 0.65 & 0.40 & 0.16 \\
\noalign{\medskip}
\hline
\multicolumn{10}{c}{Corona binaries} \\
KW & \multicolumn{9}{c}{ } \\
549 & 10.15 & 14.04 & 0.48 & 1.30 & 1.19 & 0.63 & 0.52 &      & 0.13 \\
556 & 10.55 & 14.77 & 0.55 & 1.39 & 1.12 & 0.58 & 0.52 &      & 0.85 \\
VL & \multicolumn{9}{c}{ } \\
 142 & 10.53 & 12.35 & 0.54 & 0.89 & 1.13 & 0.84 & 0.75 &      & 0.66 \\
 184 & 11.85 &       & 0.79 &      & 0.93 &      & $>$0.10 & 0.3 & 0.09 \\
1025 & 11.75 & 13.84 & 0.77 & 1.24 & 0.92 & 0.65 & 0.71 &      & 0.43 \\
Art & \multicolumn{9}{c}{ } \\
1271 & 11.34 & 11.34 & 0.71 & 0.71 & 0.97 & 0.97 & 0.99 &      &      \\
1780 & 11.22 & 11.22 & 0.69 & 0.69 & 0.99 & 0.99 & 0.99 &      &      \\
\noalign{\smallskip}
\hline
\end{tabular}
\end{flushleft}
\end{table}

\subsection{Cluster mean velocity}
We have computed Praesepe mean velocity from 46 "single" stars and found a value of
34.53 $\pm$ 0.12 (s.e.) km s$^{-1}$. The standard error is 0.81 km s$^{-1}$. The
mean value of 14 spectroscopic binaries (excluding KW 540) is 34.68 $\pm$ 0.11 km 
s$^{-1}$, with a standard error of 0.40 km s$^{-1}$. 

\begin{figure}
\resizebox{7.7cm}{!}{\includegraphics{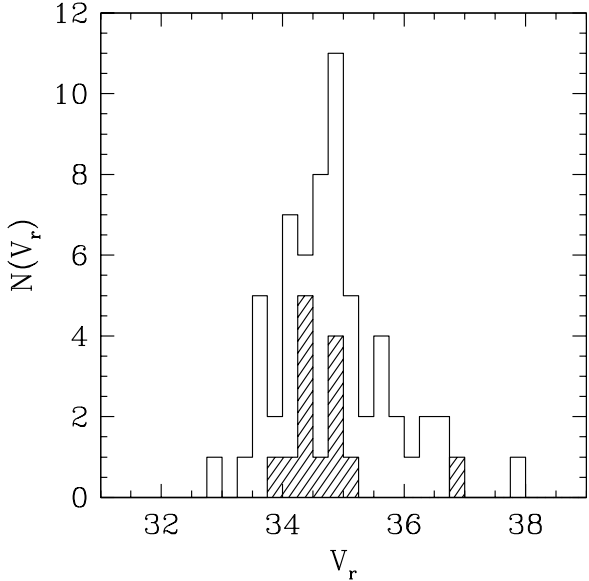}}
\caption{Radial-velocity distribution for "single" stars and spectroscopic binary 
(shaded histrogram).}
\label{meanvel}
\end{figure}

The observed velocity dispersion, 0.40 km s$^{-1}$ for the binaries and 0.81 for the 
single stars would indicate, if taken at face value, that the velocity dispersion of 
the binaries, more centrally concentrated than the single stars, is only half of
the dispersion of the single stars. This result is in agreement with what is
expected from the dynamics of star clusters. However, one should be careful and take 
into account the velocity dispersion due to unknown binaries. Fig.~\ref{meanvel}
shows the histogram of the stellar mean velocity and illustrates this effect. The 
shaded area corresponds to the spectroscopic binaries for which an orbit has been 
determined.

\begin{figure*}
\resizebox{15.0cm}{!}{\includegraphics{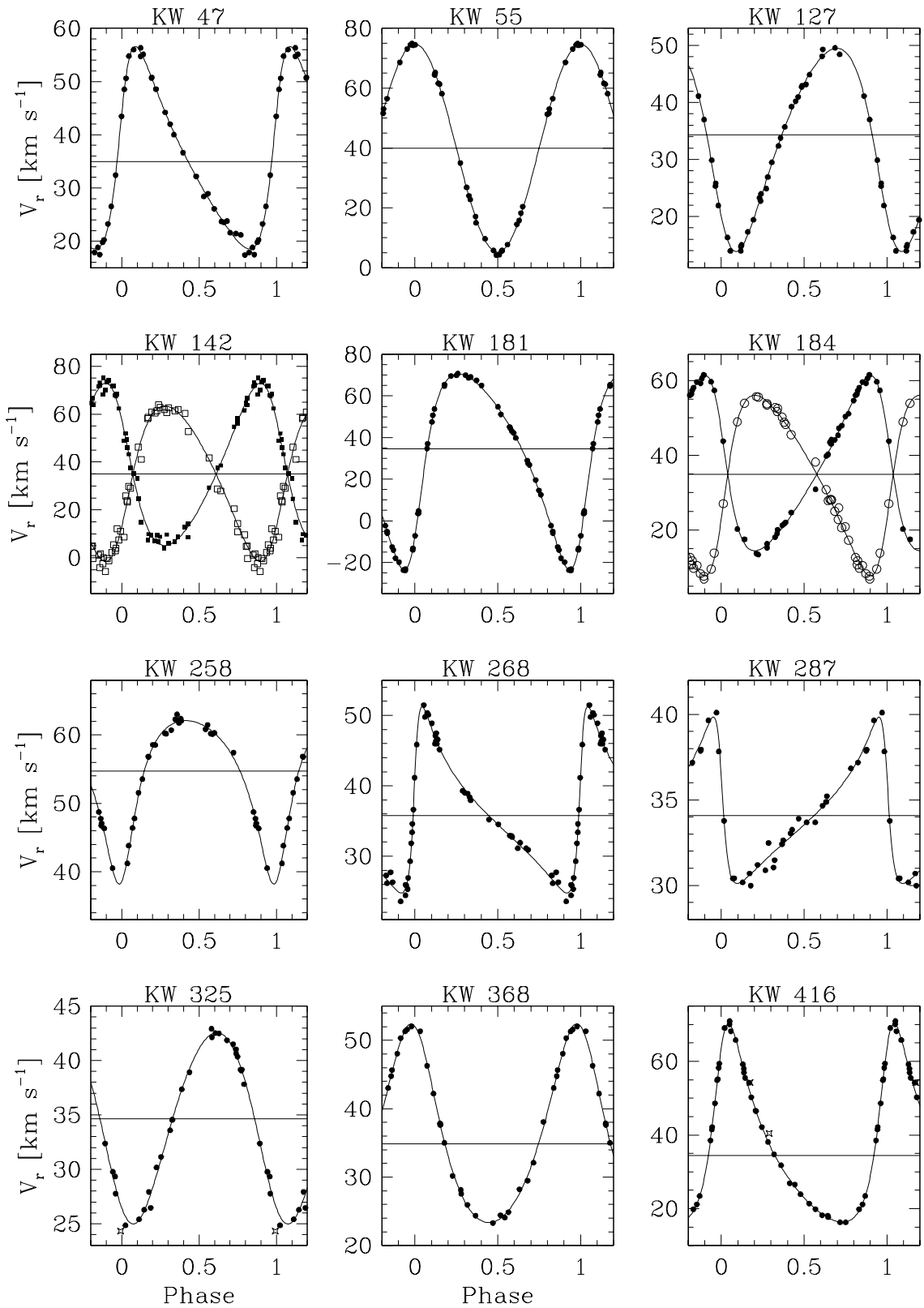}}
\caption{Radial-velocity curves for twelve spectroscopic binaries. For SB2s, filled
circles are for the primaries and open circles for the secondaries.}
\label{fig2a}
\end{figure*}

\begin{figure*}
\resizebox{15.0cm}{!}{\includegraphics{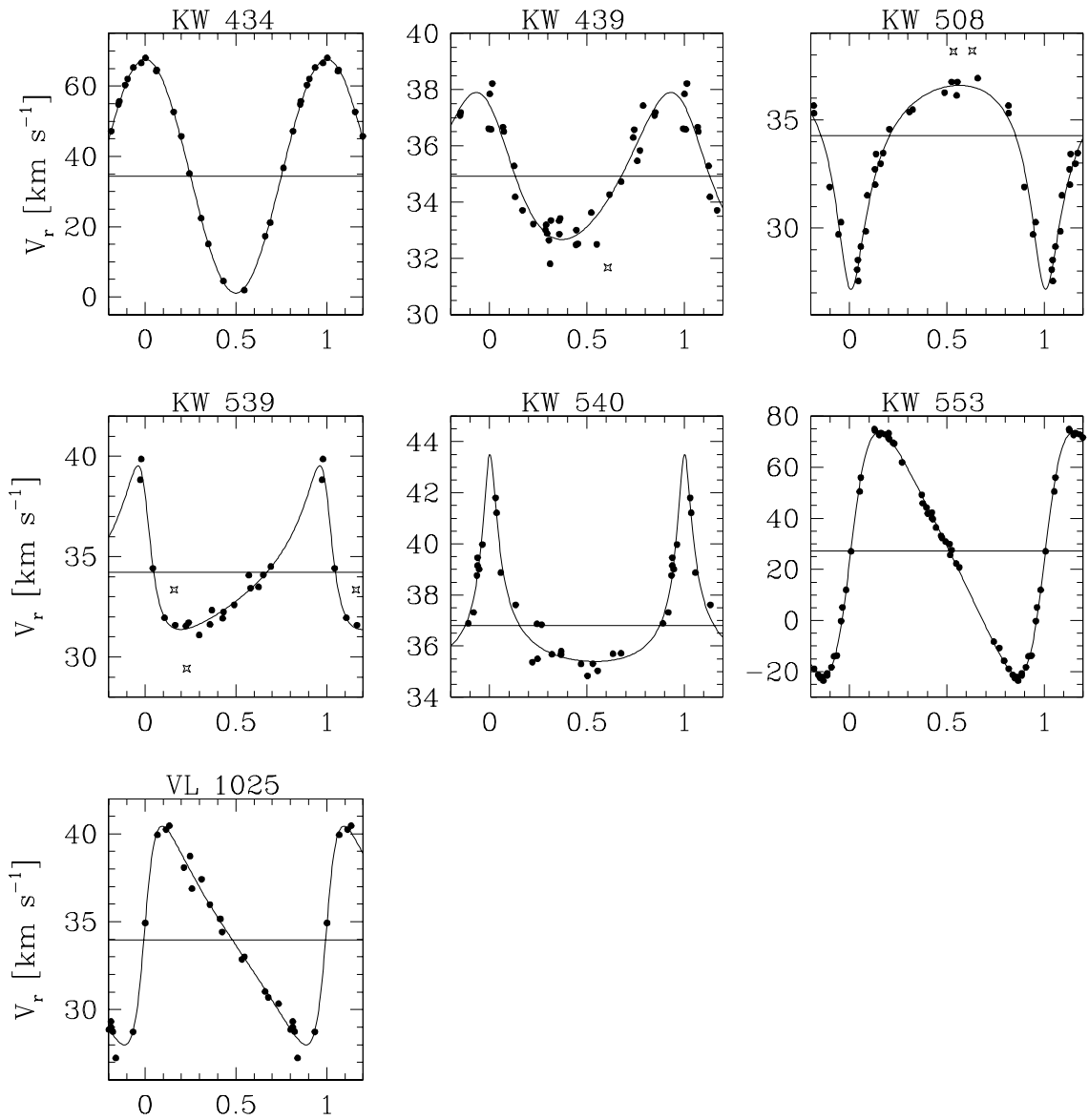}}
\caption{Radial-velocity curves for seven spectroscopic binaries. Open diamonds represent
observations rejected in the orbital solutions.}
\label{fig2b}
\end{figure*}

\section{Colour-magnitude diagram}
\begin{figure}
\resizebox{7.0cm}{!}{\includegraphics{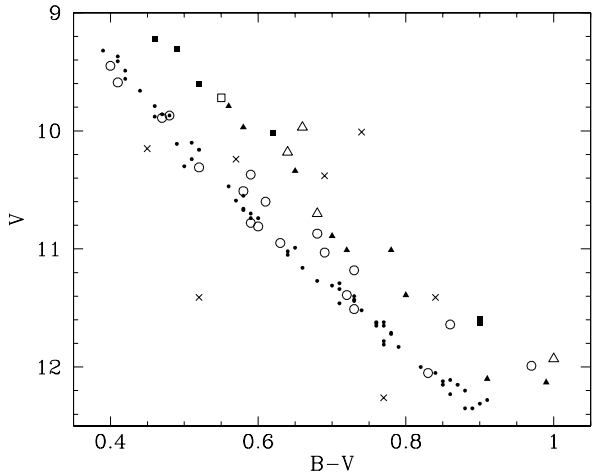}}
\caption{Colour-magnitude diagram of the F5-K0 stars in Praesepe. The solid and
dashed curves represent the ZAMS and the binary upper limit respectively. Symbols 
are as follows: points: single stars, open circles: SB1s, filled squares: SB2s,
open square: visual binary, open triangles: triple systems, filled triangles: 
photometric binaries, crosses: non-members. Notice that 10 among 17 single-lined 
binaries are located within the "single" star sequence.}
\label{cmd}
\end{figure}

The colour-magnitude diagram for the F5 - K0 stars (Fig.~\ref{cmd}) clearly shows
that binary stars are distributed all over the main sequence band. They are not
only found near the upper binary limit (0.75 mag above the ZAMS), but also among
the "single" stars. Nine binary stars are found at $\Delta V >$ 0.70 mag, 10 are found
among the single stars and 7 in the intermediate band. Therefore, any photometric
detection of binary stars in a colour-magnitude diagram will miss about 40\%
of the spectroscopic binaries.

There is however no one to one correspondance between the photometrically detected 
and the spectroscopic binaries. For (B-V) $<$ 0.90, seven obvious photometric
double stars have not been detected as spectroscopic binaries. A similar situation 
has been met in the Pleiades (Mermilliod et al.~\cite{mrdm}). Adaptive optics 
observations have permitted to resolve a number of the Pleiades photometric binaries 
(Bouvier et al.~\cite{brn}). A similar survey of 150 stars in Praesepe made in 1998 
will permit to answer this question (Bouvier et al.~\cite{bdms}).

In a similar way, it is quite possible that apparently single stars have low-mass
companions at large distance which do not produce perceptible effect on the
joint magnitudes or on the radial velocities. Data in Table~\ref{ratio} suggest that
very few binaries with mass ratios smaller than 0.3 have been detected, although the 
precision of our radial-velocity observations allows us to determine orbits with 
semi-amplitudes around 2.5 km s$^{-1}$, see for example KW 439. 

To test the presence of a variation of the number of spectroscopic binaries along the
main sequence, we have applied the test of Wilcoxon-Mann-Whitney. The test shows that, 
in our sample, binaries are more frequent among stars with small $B-V$, but the level of 
significance is 8\%. Therefore, we cannot definitely discard the possibility that this 
trend is not intrinsic, but simply comes from random noise.

In the interval 0.4 $< \,B-V\,<$ 0.55 the star distribution is clearly bimodal:
on the one hand five stars show the maximum magnitude effect: three SB2 
(\object{KW 16}, \object{KW 142}, \object{KW 496}), one visual binary (\object{KW 458}) 
and one photometric binary (\object{KW 31}). Twenty-one stars define the main-sequence locus, 
four binaries (\object{KW 47}, \object{KW 268}, \object{KW 416}, \object{KW 439}) are
photometrically indistiguishable from true single stars. 

The interval 0.55 $<\,B-V\,<$ 0.75 is much
richer in binary types. It contains the three triple systems discussed by Mermilliod 
et al. (\cite{mdm}). In \object{KW 495}, all three components are seen in the correlation
functions and the position in Fig.~\ref{cmd} is in good agreement with that expected 
for a triple system. \object{KW 365}, in which one single star and the SB primary are seen 
is very close to the 0.75 upper limit, so the contribution of the third body seems 
to be small. Then, the position of \object{KW 367}, a double-lined system with a variable 
systemic velocity is surprisingly located 0.25 mag below the upper binary limit. 
Because two components are seen, the third one should not contribute much light.
There are five single-lined binaries (KW 181, 287, 322, 325, 540) located between
0.20 and 0.50 mag above the ZAMS, with three other stars (KW 90, 334, 541) occupying
the same position, without spectroscopic detection. One (\object{KW 182}) is close to the 
upper binary limit. Finally, in this colour interval, one finds also the double-lined 
contact binary (\object{KW 244}, TX Cnc) which could be not observed with Coravel due to 
the high rotation induced by the short period.

In the redder colour interval (0.75 $< \,B-V\,<$ 0.90) the binarity is again well
marked. Six stars are at least 0.50 mag above the ZAMS: two SB2s (\object{KW 184}, \object{KW 533)}, two
photometric binaries (\object{KW 9}, \object{KW 257}) and one single-lined binary (\object{KW 297}).

In total, eleven single-lined binaries show no photometric effects of duplicity and
are completely indistinguishable from single stars. They are even often on the lower
side of the single-star locus.

\section{X-ray sources}
Because the survey of Randich et al (\cite{rs}) covers an area of 
4$^{\degr}$x4$^{\degr}$ all the Klein-Wassink area has been observed. We have
looked at the detection of the various kinds of stars: SB2s detected: 5/6, SB1s
detected: 6/17, triples detected: 2/3, photometric binaries detected: 4/8, single 
stars detected: 8/57.

From the periods determined so far, it is difficult to derive firm conclusions.
Most short periods binaries P $<$ 6$^d$ have been detected, but \object{KW 368} (P = 76$\fd$5)
has also been detected although there is no photometric evidences from the companion.
Several SB1s with much longer periods and half of the photometric binaries are
also detected. But the triple system \object{KW 365} is not. The 8 single stars detected
could also be binaries, with low-mass companions and large separation which would
explain why they have not been discovered in the radial-velocity survey.

\section{Conclusions}
The present results contribute to the recent effort to investigate the
duplicity in open clusters from various techniques and in various cluster
environments. So far only radial velocity can detect binaries with periods 
smaller than 1000 days which represents 64\% of the spectroscopic binary 
sample in Praesepe. An adaptive-optics survey of 150 G and K stars in Praesepe
will provide further information on the general duplicity in Praesepe
(Bouvier et al.~\cite{bdms}), and will allow a better comparison of the 
properties of the Hyades. The limiting separation is around 0$\farcs$1, which,
at the distance of Praesepe, 180 pc (Robichon et al. 1999), corresponds to 54 yrs for
two solar-mass stars and 76 yrs for two stars of 0.5 M$_{\odot}$. There will still
be a gap in the coverage for periods between 20 yrs and 50 yrs.
With an age of about 700 Myr, the binary properties of these two clusters are 
probably not representative any more of the primordial binary-parameter 
distribution.

The frequency of binaries in Praesepe is slightly larger than that found in the 
Pleiades (Mermilliod et al. 1992) in a similar colour interval, but the difference 
is significant at the 10\% level. Thus we cannot totally exclude that both clusters 
have similar statistical properties. Detailed comparison with the Hyades is 
awaiting the publication of the results obtained by Stefanik \&Latham (\cite{sl92}). 
But it seems that it is not possible to explain the X-ray flux differences
between the Hyades and Praesepe on the basis of difference in duplicity as 
it has been proposed by Randich \& Schmitt (\cite{rs}) and 
Barrado y Navascu\'es et al. (\cite{bsr}).

The combination of photometric and spectroscopic data results in a powerful 
tool to analyse binary stars in open clusters. For example, ten 
stars which would be classified as single on a photometric basis have been
found to be binaries from their radial velocities, and seven stars which do not 
exhibit velocity variations appear to be binary stars when their photometry is
taken into account. Thus, binary analysis is more complex than previously thought.

Based on simulations, Kroupa \& Tout's (\cite{kt}) results tend to favour a fraction 
of binaries close to unity. The present results have demonstrated the presence
of numerous binaries among the so-called single stars, but much work remains to
be done to detect companions in systems with q $<$ 0.3. Such a detection is
difficult in the visible, near-infrared photometry may help. As concerns radial 
velocities, instruments providing the precision needed to detect such low-mass 
companions are now in use, but the investment in observing time would be
rather heavy and should extend over many years. However open clusters like the 
Pleiades and Praesepe are among the best targets to study the multiplicity of 
solar-type and low-mass main-sequence stars, to build a much more complete 
distribution of mass-ratios and orbital elements.


\begin{acknowledgements}
We are grateful to Dr S. Udry for the fine reductions of the double-lines systems
and Y. Debernardi for help in computing the simultaneous solution for KW 367. We are 
also grateful to the referee, Dr J.-L. Halbwachs, for his comments and helpful discussions.
\end{acknowledgements}

\end{document}